# Micro Plasma Actuator Location Effect on Complex Fluidic Behavior: An Optimization Study

**Javad Omidi**

Research Assistant, Department of Chemical Engineering, Columbia University, NYC, USA
jo2668@colubmbia.edu

## Abstract

This study employs optimization techniques to enhance the positioning of a dielectric-barrier-discharge plasma actuator on a curved surface, taking into account diverse aerodynamic and physical scenarios. The optimization methodology utilized here is Differential Evolution (DE), complemented by an improved electrostatic model for solving electrostatic equations. In this electrostatic model, two elliptic equations, governing electrical potential and plasma density, are independently resolved, and their solutions are subsequently incorporated as source terms within the Navier-Stokes equations. Notably, contrary to prior research suggesting the placement of the plasma actuator at the leading edge of the airfoil, our findings reveal that the optimal position for plasma actuation falls within the range of 2 to 4 percent of the chord length from the leading edge, contingent upon the prevailing aerodynamic conditions. Furthermore, we elucidate a mathematical pattern in the optimization data across a continuous domain, applicable to various geometries and designs. This mathematical model articulates the optimal location as a complex function, dependent on both the linear Reynolds effect and the angular impact of the angle of attack.

## Nomenclature

| | | | |
|---|---|---|---|
| $AC$ | Alternating Current | $t_e$ | Electrode thickness, $m$ |
| $C$ | Capacitance value of DBD circuit, $F$ | $t_d$ | Dielectric thickness, $m$ |
| $CFD$ | Computational Fluid Dynamics | $u$ | Velocity component, $m/s$ |
| $DBD$ | Dielectric Barrier Discharge | $V_{app}$ | Applied voltage, $Volt$ |
| $E$ | Electric field, $N/C$ | $V_{bd}$ | Break-down voltage, $Volt$ |
| $f_b$ | Body force vector, $N/m^3$ | $x,y$ | Coordinates |
| $f$ | Frequency, $Hz$ | $\varepsilon_0$ | Vacuum permittivity, $8.854*10^{-12}\ C^2/Nm^2$ |
| $G$ | Gaussian distribution function | $\varepsilon_r$ | Relative permittivity |
| $l_e$ | Length of electrode, $m$ | $\lambda_d$ | Debye length, $m$ |
| $l_p$ | Plasma extent, $m$ | $\vartheta$ | Fluid viscosity, $m^2/s$ |
| $P$ | Pressure, $Pa$ | $\rho$ | Density, $Kg/m^3$ |
| $q_c$ | Net charge density, $C/m^3$ | $\sigma$ | Scale parameter for Gaussian function |
| $T$ | Temperature, $K$ | $\phi$ | Electric potential of external field, $Volt$ |

# 1. Introduction

Plasma actuators utilizing a Dielectric Barrier Discharge (DBD) represent a cost-effective and efficient choice for generating plasma in ambient conditions. Their affordability in both production and operation, along with attributes like compactness, lightweight construction, user-friendly operation, and absence of moving components or reliance on pneumatic or hydraulic systems, have garnered significant attention from researchers in the field of flow control devices. Recent academic studies have explored various applications of these actuators [1 – 13]. The underlying mechanisms governing these actuators are intricate, necessitating a fusion of electrical engineering principles, plasma physics, encompassing classical electric discharge physics, and flow hydrodynamics for the purpose of designing and optimizing such devices.

The widespread use of plasma actuators across various applications necessitates a deeper understanding of their optimal placement on curved surfaces. Consequently, there is a growing interest in leveraging optimization algorithms to determine the ideal actuator positions that can be applied to diverse geometries. Traditionally, most research in this area has relied heavily on experimental methods, which are not only time-consuming but also costly. Furthermore, the limitations of measurement instruments often impede the feasibility of conducting small-scale studies. However, recent advancements in numerical analysis, coupled with substantial improvements in computer processing capabilities, have paved the way for precise simulations of flow fields and flow control devices such as plasma actuators. These developments offer a promising avenue for exploring and optimizing actuator placement without the need for resource-intensive experimental models.

In the course of fundamental investigations into plasma actuators, several pivotal experiments have garnered attention. Roth et al. [14] delved into the application of plasma actuators for controlling boundary layers. Corke et al. [15] explored their utility in inducing boundary layer instability over a pointed cone at Mach 3.5, while Corke and Post [16 – 18] conducted research on static and dynamic stall control for NACA (National Advisory Committee for Aeronautics) 663-018 and NACA 0015 airfoils. Jacob et al. [19] investigated the management of laminar and turbulent flows in flat plates and low-pressure gas turbines. Orlove et al. [20] scrutinized leading-edge separation control for the NACA 0021 airfoil at various angles of attack post-stall. He et al. [21] assessed the impact of plasma actuators on flow control using the Hump model. Little et al. [22] applied plasma actuators to a NASA EET airfoil flap. Lastly, Thomas et al. [23] optimized lift coefficients through the use of a plasma actuator.

The physics underlying Dielectric Barrier Discharge (DBD) actuators is inherently intricate, involving intricate interactions among ionization, fluid dynamics, and electric fields. Achieving an accurate solution necessitates the simultaneous calculation of the Maxwell and Navier-Stokes equations. However, the computational cost associated with solving this complex nonlinear combination is generally deemed prohibitive. To address this challenge, Suzen and Huang [24] proposed an electrostatic model for electromagnetic plasma, drawing on Enole's experiment [25] as a basis. Their approach involved simplifying Maxwell's equations into two elliptic equations by assuming the formation of quasi-steady plasmas and neglecting magnetic forces. The Lorentz body force was utilized as a source expression, applied to the fluid flow equations. Additionally, for distributing the load on the dielectric surface, a one-dimensional Gaussian distribution was employed, grounded in experimental data [26]. Various modifications to this model have been put forth, including contributions by Bouchmal [27], Skote et al. [28], Abdullahzadeh et al. [29], and the author [30 – 33]. The author extended Suzen and Huang's model [24] by amalgamating several numerical models, thereby facilitating the analysis of how voltage and frequency impact actuator performance [30]. Furthermore, the author introduced an improved phenomenological electrostatic model, independent of experiments, which is also employed in this study [31].

The best location in actuator installation is an important factor in determining the optimal usage of plasma actuators. Jolibois et al. [34] employed a DBD actuator to regulate the separation current on the NACA (National Advisory Committee for Aeronautics) 0015 airfoil in this regard. The ultimate purpose of their experiments was to better understand where to actuate along the chord to be most successful (as a function of angle of attack). These investigations suggest that the plasma actuator is most effective when it works near to the natural separation point. Bormel et al. [35] studied the influence of the plasma actuator on the lift and drag coefficients generated by the flow over the NACA (National Advisory Committee for Aeronautics) 4415 airfoil using particle image accelerometer (PIV) observations. DBD actuators were placed at the leading edge, 30%, and 60% of the chord from the leading edge. It was discovered that in order to improve lift or drag, the actuators needed to be positioned closer to the leading edge and in front of the separation starting point. Salmasi et al. [36] investigated the effects of plasma actuator location on fluid flow passing the NLF0414 airfoil using numerical and experimental methods. The location of the plasma actuator on the airfoil was modified in this example, and the effect on the separation delay point was explored. The results demonstrate that the actuator has the greatest effect on delaying separation when it is placed directly on the leading edge of the airfoil, with about 100 percent improvement in airfoil efficiency.

Our significant contribution in this study is the development of an optimization technique for determining the best location for actuator installation on a controlled aerodynamic curved surface under varied aerodynamic conditions. Furthermore, we formulate this approach for a given range of Reynolds number and angle of attack for a basic airfoil as a benchmark in this work. This finding could be applied to any geometry or fluid flow condition in further studies.

In Section 2, the problem is introduced, and in Section 3, the principal governing equations for the fluid flow and electrostatic, as well as the optimization algorithm, are presented. The authors' proposed improved boundary conditions and parameters are also discussed in this section. Section 4 discusses EHD and optimization solvers validation. The optimization findings are presented in two phases in Section 5 and the advances and limitations of the applied scheme is also considered here, followed by the final conclusion.

## 2. Problem Description

After conducting an extensive investigation aimed at enhancing the performance of a phenomenological model for simulating the impact of plasma actuators on fluid flow, as detailed in a prior study [31], and considering its ability to account for the influence of geometric [32] and operational [33] parameters, this work introduces a Differential Evolution (DE) optimization algorithm. The goal is to identify the optimal placement of a plasma actuator on a curved surface to achieve superior performance across various angles of attack and Reynolds numbers. For this purpose, the NACA 0015 airfoil, a well-established airfoil design, was selected for implementation. To achieve the highest aerodynamic performance of the airfoil, the plasma actuator can be positioned at various locations, each potentially optimal for different flow conditions. Based on the insights from previous research in plasma actuator design, the following geometric characteristics are considered in the current study: a horizontal gap between electrodes equal to 0.002 times the airfoil chord length (C), an embedded electrode length of 0.04C, an exposed electrode length of 0.032C, a dielectric thickness of 0.006C, and an electrode thickness of 0.00015C. The chosen reference material possesses a relative permittivity of 3. Furthermore, the plasma actuator's operating settings are consistently configured to 12 kV voltage and 8 kHz frequency.

The quest for the optimal placement of the plasma actuator involves a two-stage approach. Given that the primary objective of this optimization is to forestall flow separation, it is advantageous to initially consider installing the actuator within the first 25% of the chord length from the leading edge. This choice is informed by the observation that under complete stall conditions, the separation of the NACA 0015 airfoil extends all the way to the leading edge [34]. Consequently, in the first phase of the study, a parametric investigation is employed to identify the most favorable range of positions for the plasma actuator within the 0 to 25% region, specifically for a Reynolds number of 600k. Subsequently, the findings from this initial study serve as the basis for conducting optimization calculations utilizing the Differential Evolution (DE) algorithm.

In outlining an optimization problem, we can identify three fundamental elements: the objective function, design variables, and constraints. In this context, the objective function aims to minimize the airfoil's drag-to-lift coefficient ratio. The optimization variables encompass plasma actuator positions spanning from 0 to 25% of the chord length measured from the leading edge, as well as angle of attack values constrained within 18 to 26 degrees, and Reynolds numbers ranging from 300k to 900k. Here we have assumed the mutation factor $F=0.5$, and $C_r=0.8$, $N_p=5$. If convergence is achieved at generation 50, it requires 250 times solution of the electrostatic and hydrodynamic equations in our grid.

## 3. Governing Equations

### 3.1. Electrostatic Model

When a high Alternating Current voltage is applied to a configuration comprising two electrodes separated by a dielectric layer, the adjacent air undergoes ionization in an intermittent manner, giving rise to a small plasma region. The presence of ionized particles in an electric field results in the generation of a body force that acts on the flow in a quasi-steady manner. This body force can be incorporated into the Navier-Stokes equations as a source term. Among the various approximations available, the Suzen-Huang (S-H) model [24] stands out as one of the more physically grounded methods [30] for calculating this generated body force. In this model, the body force components are determined by solving two elliptic equations that are independent of the fluid flow equations. The Lorentz equation is employed to describe the body force components, irrespective of the influence of magnetic forces.

$$\vec{f_b} = q_c \vec{E} \qquad (1)$$

where, $\vec{f_b}$ is the body force vector per unit volume, $q_c$ is the charge density in C/m$^3$, and $\vec{E}$ is the electric field vector. Suzen and Huang [24] made several key assumptions to simplify their model. They assumed that ionized particles have sufficient time to redistribute, rendering the ionization process quasi-steady. They proposed that the electric potential can be considered as a combination of two components: the potential generated by the external electric field and the potential arising from the net charge density. This assumption stems from the fact that gas particles remain weakly ionized during the plasma formation process. Furthermore, they hypothesized that the Debye thickness is small and that the charge on the wall above the enclosed electrode is minimal. Consequently, the distribution of charged particles within the domain is primarily determined by the potential created by the electric charge on the wall, which remains constant even when the external electric field changes. To account for these assumptions, they utilized two distinct equations to describe the distributions of the electric potential field and the charge density, as dictated by Maxwell's equations.

$$\nabla . (\varepsilon_r \nabla \phi) = 0 \qquad (2)$$
$$\nabla . (\varepsilon_r \nabla q_c) = \frac{q_c}{\lambda_d^2} \qquad (3)$$

where $\phi$ is the electric potential, $\lambda_d$ is the Debye length, and $\varepsilon_r$ is the relative permittivity.

## 3.2. Modified Boundary Conditions

To solve two differential Equations (2) and (3), two different sets of boundary conditions must be used. In both the dielectric and fluid domains, the electric potential Equation (2) is solved as shown in Figure 1. The following are the definitions of boundary conditions: $\partial \phi / \partial n_i = 0$ on outer boundaries, $\phi = \phi(t)$ on the exposed electrode, and $\phi = 0$ on the embedded electrode. Note, $n_i$ is the unit vector normal to the surface and $\phi(t) = \phi^{max} f(t)$ denotes the applied voltage. Also, $\phi^{max}$ refers to the amplitude of the applied AC voltage. The wave form function $f(t)$ is a time-dependent function that works for both steady and unsteady actuators.

Only the air side of the domain is solved for the net charge density Equation (3) which could be seen in Figure 1. The following are the boundary conditions for solving this equation: $q_c = 0$ on outer boundaries, $q_c = q_c^{max} G(x) f(t)$ downstream of the exposed electrode and above the embedded electrode on the dielectric surface, named charge surface, and $\partial q_c / \partial n_i = 0$ on the solid walls, with the exception of the lower electrode's zone. $q_c^{max}$ refers to the maximum charge density of the applied AC voltage on the dielectric surface. Based on the experimental results [25], in S-H model [24] a half Gaussian distribution $G(x) = exp(-\frac{\tilde{x}^2}{2\sigma^2})$ was applied to calculate the variation of the charge density on the charge surface. $\tilde{x}$ is the chord-wise length measured from the leading edge of the embedded electrode, and $\sigma$ is a scale parameter for the Gaussian distribution. $\lambda_d$ and $q_c^{max}$ remain to be determined later by an empirical or phenomenological model.

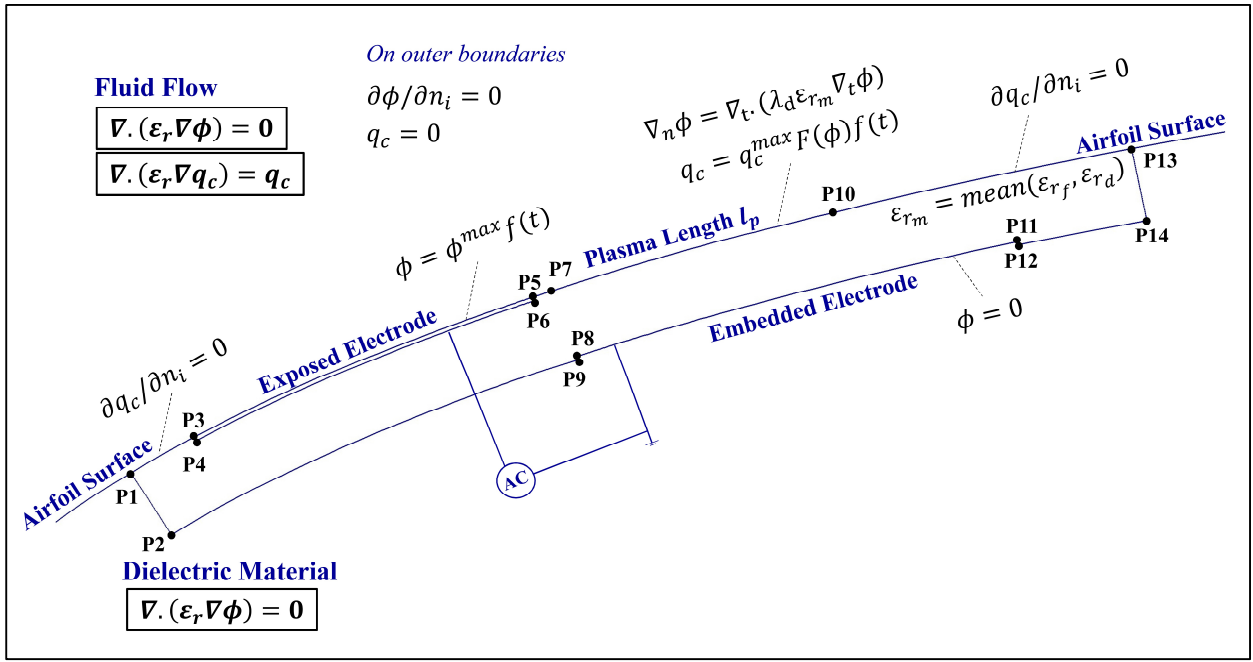

**Figure 1. Boundary conditions for electrical potential and density charge on a schematic figure of plasma actuator**

One limitation of the Suzen-Huang (S-H) model is its assumption that the charge density remains independent of the increasing applied voltage. This limitation arises from the decoupling of the governing equations, which restricts the model's applicability to specific geometries and scenarios. Moreover, the S-H model's representation of the wall jet is excessively thin [24], and the resultant velocity profiles of the induced jet do not align with experimental findings [31]. Consequently, the model lacks accuracy in the hydrodynamic field, particularly in the critical boundary layer region. In an effort to address these limitations, Ibrahim and Skote [28] enhanced the boundary conditions of the S-H model, leading to more accurate solutions for the electric potential distribution. They introduced the following boundary condition for the governing equation of the electric potential field:

$$\nabla_n \phi = \nabla_t \cdot (\lambda_d \varepsilon_r \nabla_t \phi) \qquad (4)$$

where $\nabla_n$ and $\nabla_t$ describe the normal and tangential derivatives in respect to the surface.

In response to the shortcomings of the S-H model, we have introduced an improved boundary condition for the charged surface, replacing the Gaussian distribution utilized in our prior publications [31]. Under this new boundary condition, the charge density on the electrode boundary surface is directly proportional to the electric potential on the electrode surface. This value is then applied in Equations (5) to establish the boundary condition for the net charge density Equation (3) along the plasma extent. This process is implemented subsequent to solving the electric potential Equation (2) and ascertaining the distribution of the potential electric field on the charged surface:

$$0 < x < 17\% \qquad q_c(x) = q_c^{max} \left( \frac{\phi_{max}^{local} - \phi}{\phi_{max}^{local} - \phi_{17\%}} \right)^{2.0}$$

$$17\% < x < 100\% \qquad q_c(x) = q_c^{max} \left( \frac{\phi - \phi_{min}^{local}}{\phi_{17\%} - \phi_{min}^{local}} \right)^{0.6} \qquad (5)$$

### 3.3. Semi-Empirical Parameters Involved in the Model

One might anticipate that the thrust produced by the plasma actuator is directly related to the electric energy consumed by the DBD actuator [31]. Assessing the energy consumption of the AC circuit provides a highly accurate approach for estimating the performance of the plasma actuator. Yoon and Han [37] devised a model representing the dielectric barrier plasma actuator as an alternating current circuit consisting of two capacitors connected in series. They computed the thrust value by considering the energy consumed by these capacitors. $C_d$ is the capacitance of the lower electrode and the dielectric barrier, while $C_g$ is the capacitance of the upper electrode and the generated plasma over the charge surface. The capacitance of the capacitors can be expressed in terms of the geometric and material properties of the actuator components as follows:

$$C_g = 2\pi\varepsilon_0 \frac{l_p}{\ln\left(\frac{0.5t_e + \lambda_d}{0.5t_e}\right)} \qquad (6)$$

$$C_d = 2\pi\varepsilon_d \frac{l_e}{\ln\left(\frac{0.5t_e + 2t_d}{0.5t_e}\right)} \qquad (7)$$

where, $t_e$ is the thickness of the electrode, $t_d$ is the thickness of the dielectric barrier, $l_p$ is the length of the plasma extent on the charge surface, $l_e$ is the length of the embedded electrode, $\varepsilon_0$ is the permittivity of the free space and $\varepsilon_d$ is the dielectric permittivity.

Bouchmal [27] addressed an inverse problem by employing the S-H model along with experimental data from Kotsonis et al. [38] regarding body forces. The objective was to determine the Debye length along the charged surface, and this information pertains to a 2 kHz frequency and a range of applied voltages. The approach involved utilizing the Laplace Equation (2) and assuming that the charge density at each segment within the flow domain was unknown. By using the known body force at each segment, the charge densities across the entire flow domain could be calculated, leading to the charge density distribution over the charged surface. Subsequently, the Debye length was determined as the height of the evolved charge in the solution domain. Consequently, Bouchmal [27] regarded the Debye length as a linear function of the applied voltage:

$$\lambda_d[m] = 0.2(0.3 \times 10^{-3} V_{app} - 7.42 \times 10^{-4}) \tag{8}$$

where, $V_{app}$ is the applied voltage in kV. In this equation, the dependence of the Debye length on frequency is neglected.

The Debye length is typically influenced by both the applied voltage and frequency, as reported by Kotsonis et al. [38], although its dependency on frequency becomes less significant at higher frequencies [33]. During our inverse analysis to extract the charge distribution from body forces, we observed that the Debye length exhibits a frequency dependence. Equation (9), employed as a correction factor for Equation (8), is derived by fitting a curve to these experimental data [31].

$$\alpha_{\lambda_d}(f) = 0.5611 Arctan(-170.3(f)^{-5.124}) + 1.768 \tag{9}$$

The frequency in this equation is given in kilohertz (kHz). To ensure the Debye length remains unaffected by high frequencies [31], a tangent inverse function is incorporated in this curve-fitting process. This adjustment is made to accurately simulate the asymptotic effect of frequency on the Debye length. This modified equation effectively predicts the distribution of charge density along the dielectric surface, the resulting body force, and the velocity profiles within the boundary layer. Furthermore, this model helps circumvent the compatibility issues explored by Suzan and Huang [24].

### 3.4. The Non-Dimensionalized Form and the Numerical Simulation Process

Non-dimensionalizing the equations and boundary conditions while solving the electric potential and charge density equations is a preferred approach. Although Equations (2) and (3) are time-independent, the boundary conditions at the exposed electrode in Equation (2) and the charging surface in Equation (3) are time-dependent. We can eliminate the time dependency of the applied voltage using our non-dimensionalization method, as we assume that this time variation is distinct from the hydrodynamic characteristics of the domain. The normalized parameters are defined using Equations (10). By applying this normalization technique to the two governing equations of the electrostatic model, we obtain the non-dimensionalized Equations (11) and (12).

$$\phi^* = \phi/\phi^{max} f(t)$$

$$q_c^* = q_c/q_c^{max} f(t)$$

$$\vec{E}^* = \frac{\vec{E}}{E_0} = l_p \nabla \phi^* \tag{10}$$

$$E_0 = \frac{V_{app} - V_{bd}}{l_p}$$

$$\nabla \cdot (\varepsilon_r \nabla \phi^*) = 0 \tag{11}$$

$$\nabla.(\varepsilon_r \nabla q_c^*) = \frac{q_c^*}{\lambda_d^2} \tag{12}$$

In simulating fluid flow, two-dimensional incompressible Reynolds-Averaged Navier-Stokes (RANS) equations are employed [30]. Since most of the energy provided to the plasma actuator goes toward accelerating the fluid particles, with only a minor portion contributing to fluid heating [30], the energy equation for the flow field is omitted. The primary equations for mass and momentum conservation that are utilized to simulate fluid flow are as follows:

$$\nabla.\vec{u} = 0 \tag{13}$$

$$(\vec{u}.\nabla)\vec{u} = -\frac{1}{\rho}\nabla P + \upsilon \nabla^2 \vec{u} + \vec{f_b} \tag{14}$$

in which $\vec{f_b}$ is the body force per unit volume in N/m$^3$ due to the effect of plasma actuator. $\vec{u}$, $\rho$, $P$ and $\upsilon$ are the velocity, the density, the static pressure, and the kinematic viscosity, respectively. The body force generated by the plasma actuator is added to the right hand side of the momentum equation, as shown in Equation (14).

### 3.5. Differential Evolution (DE) Optimization

Storn and Price [39] were pioneers in utilizing the differential evolution optimization algorithm to fit a polynomial to the Chebychev function. This non-gradient optimization approach has demonstrated remarkable effectiveness in identifying extrema in continuous spaces, making it a commonly used technique in scientific and engineering applications. Unlike gradient-based optimization algorithms that calculate a gradient to determine the steepest slope direction in the objective function domain, this algorithm is akin to both random search (genetics-based) approaches. Instead, it relies on vector subtraction to pinpoint extrema, effectively emulating the gradient computation. This model has proven its ability to swiftly and accurately locate global extrema.

The DE optimization process comprises four fundamental phases: 1) Initialization: In this phase, the first generation is generated randomly. Each generation consists of a predefined population size, with each individual characterized by its properties, typically represented as genes or design variables in a vector. 2) Mutation: This step involves introducing changes to the population, such as restoring lost or unknown genes. This helps prevent the optimization process from converging to local extrema. 3) Recombination: During this phase, the current population is combined with the mutant population, resulting in random structural interactions. This step allows for the development of improved individuals by leveraging the knowledge from the previous best individuals. 4) Selection: In the final phase, individuals are selected based on natural selection principles. The best-performing individuals from the current population and the recombined individuals are retained, while weaker solutions are discarded. This process ensures the progression of the optimization toward more favorable solutions.

### 3.6. The Optimization Algorithm

Initially, we establish a minimization objective function, which is dependent on D design variables. Typically, the population size of the first generation is set at 5 to 10 times the number of design variables D. However, in cases like ours where evaluating the objective function demands a considerable amount of time, a smaller population size is often preferred. Each member $i$ of the generation $G$ is presented by a vector $x_{i,G}$ in a D-dimension space (Equation (15)). Each entry of this vector is one attribute (gen) of this member, and each design variable is limited to its lower and upper limits, by Equation (16).

$$x_{i,G} = [x_{i,G}^1, x_{i,G}^2, \dots, x_{i,G}^D], \quad i = 1, 2, \dots, N_p \tag{15}$$

$$x_{min}^j \leq x_{i,1}^j \leq x_{max}^j, \quad j = 1, 2, \dots, D \tag{16}$$

In the mutation step to generate a new member corresponding to $x_{i,G}$, three different members of the current population are randomly selected, namely, $x_{r_1,G}$، $x_{r_2,G}$ و $x_{r_3,G}$, where $i$, $r_1$, $r_2$, $r_3$ are different numbers. According to Equation (17), a new vector is made up of these three, i.e.

$$\vartheta_{i,\ G+1} = x_{r_1,G} + F(x_{r_2,G} - x_{r_3,G}) \tag{17}$$

where F is the mutation factor, which defines the magnitude of our evolutionary direction $(x_{r_2,G} - x_{r_3,G})$. A value between 0 and 2 is commonly chosen for $F$. A tiny $F$ is more ideal for finding the absolute extremum if there are several closed extremums, but it is excessively time demanding. In order to ensure that computations remain comprehensible when dealing with small mutation factors, a denser population becomes essential. In many aerodynamic design challenges, a mutation factor ranging between 0.4 and 1 is typically selected.

In the next step, recombination, a test vector $u_{i,G+1}$ is generated, whose components are a random selection of components of $\vartheta_{i,G+1}$, and $x_{i,G}$ as shown in Equation (18). This random selection is based on a number $C_r$, which ranges from 0 to 1, and affects how much a new member inherits qualities from its parent $x_{i,G}$ or other previous generation members. As a result, a random number between 0 and 1 is generated for each design variable (dimension) $j$, and if it is less than $C_r$, other members' attributes are inherited instead of its parent's. To guarantee that the old member is not reselected, we also generate an integer random number $I_{rand}$ between 1 and $D$, and at least the $j^{th}$ attributed is inherited from other members ($u_{i,G+1} \neq x_{i,G}$ و $\vartheta_{i,G+1} \neq u_{i,G+1}$).

$$u_{i,\ G+1}^j = \begin{cases} \vartheta_{i,G+1}^j & if \quad Rand_j[0,1) \leq C_r \quad or \quad j = I_{rand} \\ x_{i,G+1}^j & if \quad Rand_j[0,1) > C_r \quad and \quad j \neq I_{rand} \end{cases} \tag{18}$$

The concluding step involves comparing the target (parent) and test vector (as represented in Equation (19)) to ascertain which one exhibits the lowest (or highest) value of the objective function.

$$x_{i,G+1} = \begin{cases} u_{i,G+1} & f(u_{i,\ G+1}) \leq f(x_{i,G}) \\ x_{i,\ G} & otherwise \end{cases} \tag{19}$$

### 3.7. Geometrical Modeling and Grid Generation

The vertices of the exposed electrode are precisely defined by specific points located at each corner of the electrodes and the plasma extensions on the surface, as illustrated in Figure 1. These points, which are perpendicular to the surface, can be computed numerically by utilizing the surface points of the airfoil and the derivative of the surface function. In this context, the surface function for the NACA 0015 airfoil is represented by Equation (20), and the derivative of this function is employed to determine the perpendicular points. These 14 values, as depicted in Figure 1, are adequate for identifying all such locations at each stage of the optimization process when designing the geometry of the DBD actuator. Additionally, various sub-points essential for defining the dielectric solution domain, including midpoints employed to align the electrode surfaces parallel to the primary airfoil surface, have been integrated into the production process.

$$y = \pm 0.75[0.2969\sqrt{x} - 0.1260x - 0.3516x^2 + 0.2843x^3 - 0.1015x^4] \tag{20}$$

Considering the low-velocity nature of the fluid flow and the relatively high Reynolds number, the flow remains incompressible, with the transition to turbulence and separation occurring predominantly on the suction side of the airfoil. Managing these phenomena necessitates adhering to specific domain and grid requirements. In order to assess the amplitude of the plasma and its interactions with the flow field, the grid generated must undergo modifications around the electrodes. To account for the impact of the wake zone on the flow field, the computational domain extends 20 chord lengths upstream, as well as upwards and downwards along the airfoil, and 40 chord lengths downstream. A grid study was conducted to ensure the results' grid independence. After evaluating the generated grids, a C-type grid [24] comprising multiple blocks and a total of 60,300 cells was chosen. This grid configuration ensures that the maximum and minimum $Y^+$ values for the SST Transition model are 0.83 and 0.33, respectively, with an average of 0.58, which is well-suited for this study [40].

### 3.8. The Optimization Process and Flowchart

Figure 2 outlines the flowchart of the optimization algorithm, which comprises three crucial steps: 1) Generation of Initial Candidates: In the first step, NP candidates are randomly generated, adhering to the constraints defined in Equation (16). Each candidate represents a unique location for the DBD plasma actuator. Each location comes with its own set of design variables, which are selected based on the optimization constraints. 2) Grid Regeneration and Computation: The second stage involves regenerating the grid within the airfoil's dielectric material, influenced by the geometrical model of the actuator on the airfoil. Subsequently, the electrostatic model is employed to compute the charge density and electrical potential. Objective functions for each candidate in the population are derived by solving the Navier-Stokes equations. 3) Evaluation and Generation of New Candidates: In the third phase, the optimization algorithm evaluates the results. A new generation of candidates is formed based on the optimization constraints. These last two stages are iteratively repeated until the solution converges to a final optimized solution.

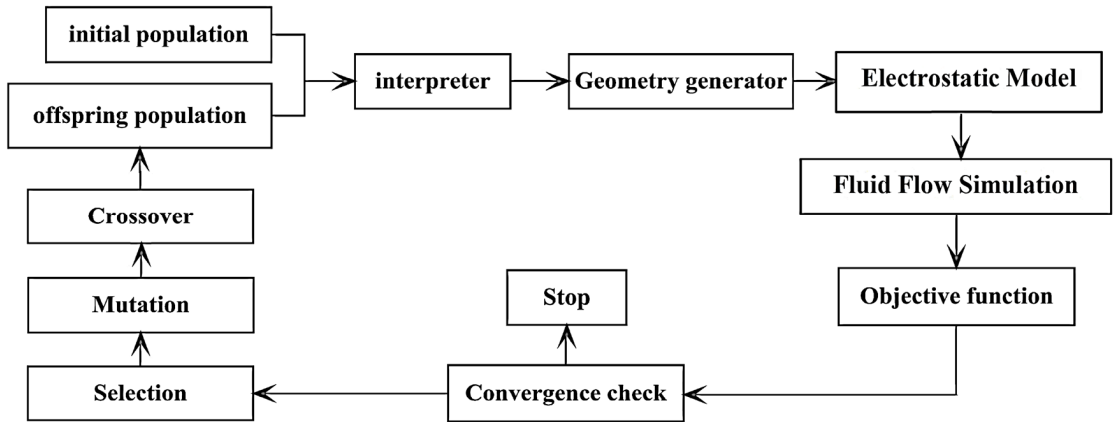

**Figure 2. Flowchart of the optimization algorithm**

## 4. Solver Validation

### 4.1. DE Optimization

We utilize the absolute extremum of the well-known Schwefel function with several local extremums to validate our optimization code:

$$f(x) = \sum_{i=1}^{n} -x_i Sin\left(\sqrt{|x_i|}\right) \qquad -500 \leq x_i \leq 500 \qquad (21)$$

For a two dimensional space, n=2, the absolute minimum is located at $x_i = 420.968$. We used a population of 25, with a mutation factor $F = 1.5$, with $C_r = 0.5$. Distribution of the population at six stages are shown in Figure 3. As one observes, all population was converged to the absolute minimum in the 300$^{th}$ generation, while the extremum was found in the 120$^{th}$ generation.

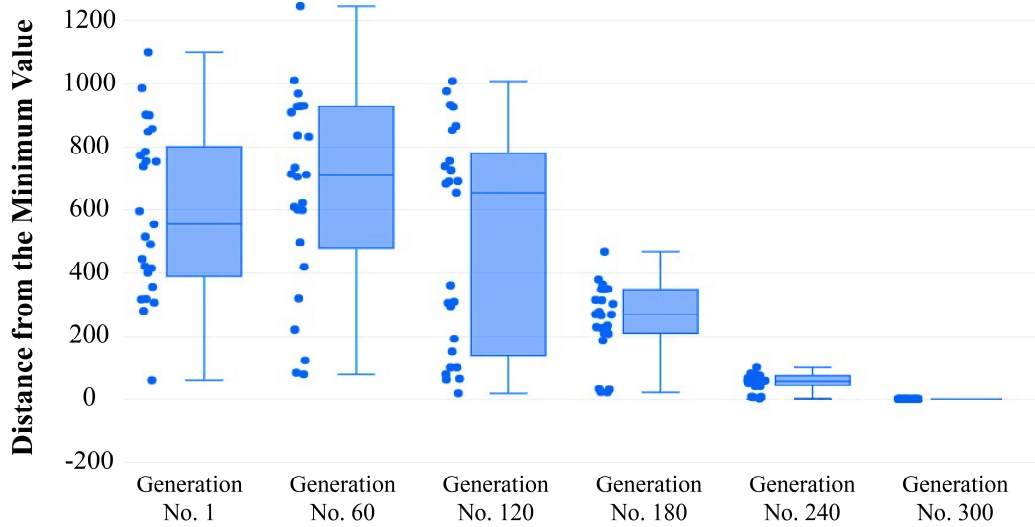

Figure 3. Using differential evolution to find the minimum of the objective function with $N_p$=25, F=1.5, $C_r$=0.5, for various generation numbers

### 4.2. EHD Solver

To validate the accuracy of the improved electrostatic model in simulation of the interaction of the actuator induced flow with the boundary layer flow, the PIV data of reference [41] describing a stagnant flow over a flat plate is used. Figure 4 compares the boundary layer profile at four different locations. Here the applied voltage is 8.8 kV. The external and internal electrodes' lengths are respectively 2.5 mm and 10 mm, and the dielectric thickness is 0.4 mm, made of Kapton.

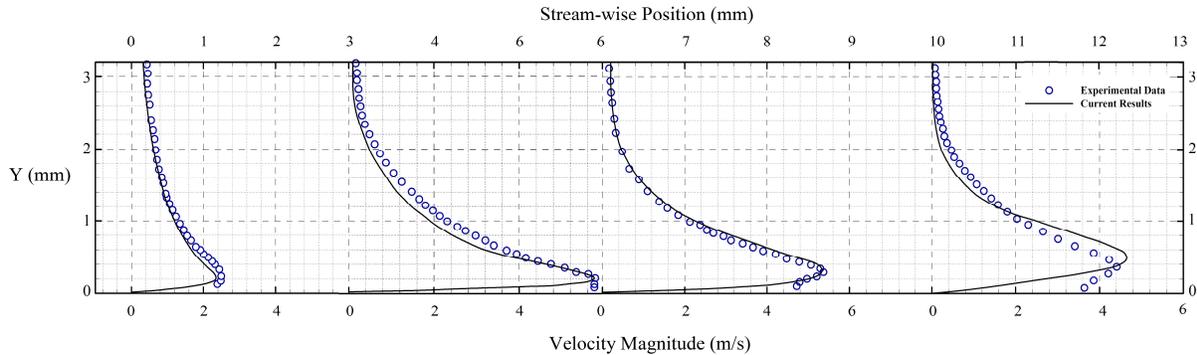

Figure 4. Comparison of the boundary layer velocity profile over a flat plate of the current simulation with the PIV data [41]

## 5. Results and Discussion

### 5.1. Optimization Phase 1

In the first phase of optimization, the best range of position for the plasma actuator is determined, and then the optimization algorithm is used in the second phase. After a complete stall, the control of flow separation at various angles of attack (AOA) is examined for this phase. The actuator is installed in six different positions at the same distances in this parametric analysis. The plasma actuator should be installed between the leading edge (LE) and 25% of the chord length from the LE at these 6 points. This is due to the fact that with thin airfoils like NACA (National Advisory Committee for Aeronautics) 0015, after complete stall, the boundary layer separation develops to the LE [34]. As a result, momentum injection by the actuator should be concentrated in these places to avoid the boundary layer from separating and the formation of undesired pressure gradients on it. These six places are at zero, five, ten, fifteen, twenty-five, and twenty-five percent of the chord length (C) from the LE. This study was done for a Re of 500k.

Figure 5 shows the performance of the plasma actuators in terms of lift coefficient when they are mounted in the six locations. The actuator effect has only been studied at AOAs near to the stall point of the airfoil due to the insufficient influence of the actuator before the stall angle [30]. As can be seen, the clean airfoil is at its maximum lift coefficient at an AOA close to 16.4 degrees and at AOAs of 20 to 22 degrees, the separation has progresses entirely to the areas near the leading edge. In this condition, there is a sharp decline in the lift coefficient.

At the angles where the flow separation occurred, Figure 5 indicates an increase in the lift coefficient for the actuated airfoils. Among the other locations, the airfoil with a 5% actuator had the best performance. Contrary to predictions [34 and 35], installing the actuator on the LE had less of an effect than putting it at 5%. The lift coefficient diagram, however, shows that at an angle of 26 degrees, the airfoil with the actuator on the LE performed very close to the airfoil with the actuator at 5%.

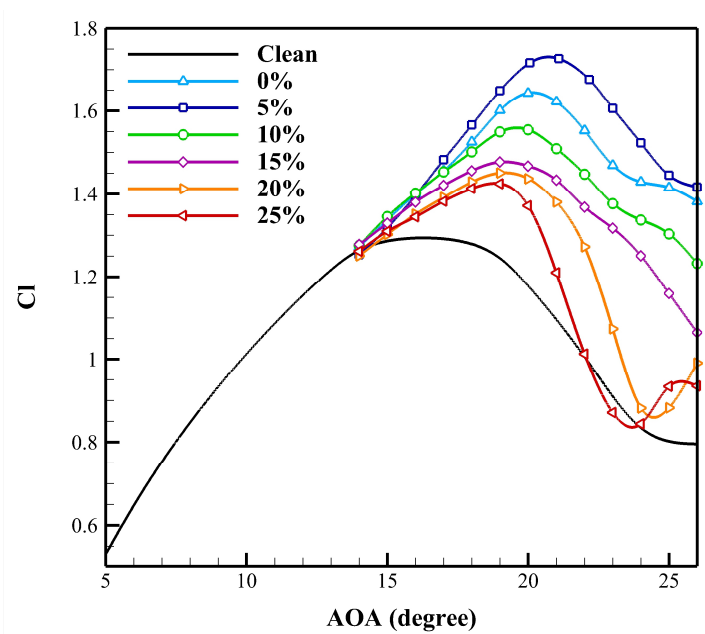

**Figure 5. The effect of plasma actuator position on the lift coefficient at different angle of attack**

The airfoils with actuations at the 20 percent and 25 percent positions have undergone static stall earlier than the others, as seen in Figure 5. The reason for this is that the separation vortex includes the actuator region at 20 and 25 percent. As a result, at an AOA of 22 degrees, these actuators are totally placed in the flow separation area, and their influence on enhancing the flow is gone. Even at this AOA, the performance of the airfoil with a 25% actuator is worse than that of the clean airfoil.

At different AOAs, the airfoil with an actuator in the 15 percent position exhibits irregular performance. Figure 5 illustrates the loss of airfoil efficiency at high AOAs for actuators located a distance from the LE, since the placing of the actuator in the separation region causes a type of vortex in the reverse flow, which is undesired.

The percentage of the maximum lift coefficient enhancement produced by the actuator is provided in Figure 6 for a more extensive investigation of the effect of utilizing plasma actuators in different locations on the airfoil. Also, Table 1 provided data of stall delay and maximum lift coefficients for various actuator installments. As previously stated, utilizing the actuator in the 5% position results in the greatest improvement in the lift coefficient. The maximum value of the lift coefficient increases by 33.59% when the actuator is in this position, as shown in Figure 6. Following that, for actuators positioned on the LE, a 26.83% increase in the maximum lift coefficient of the airfoil is visible. Similarly, airfoils with actuators put in 10%, 15%, 20%, and 25% positions achieved coefficient increases of 20.46%, 14.09%, 11.97%, and 10.23%, respectively. Figure 6 enhances the probability of obtaining an optimum location in the area between the LE and a position 5% of the chord length from the LE for plasma actuator installation and momentum injection with a better functional position.

Table 1. The maximum lift and delay in airfoil stall caused by actuator location

| Actuator Location | Stall Delay (angle) | Maximum Cl |
|---|---|---|
| 25% | 2.29 | 1.43 |
| 20% | 2.84 | 1.45 |
| 15% | 2.95 | 1.48 |
| 10% | 3.32 | 1.56 |
| 5% | 4.39 | 1.73 |
| 0% | 3.76 | 1.64 |

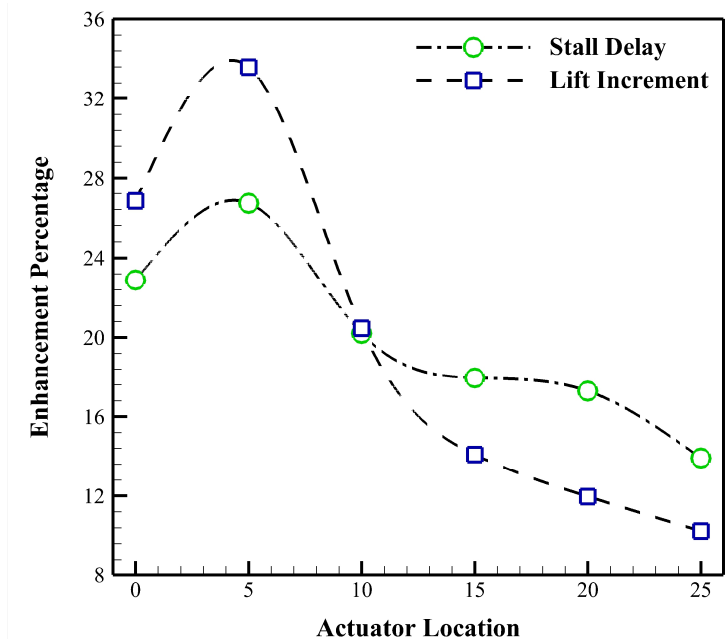

Figure 6. The lift increment and delay in airfoil stall caused by actuator position

Figure 6 also shows the percentage of the delay in the actuated airfoils' stall angles relative to the clean airfoil. As predicted, injecting momentum into the boundary layer to regulate flow separation causes a delay in the stall angle proportional to the rise in the maximum lift coefficient. Figure 6 depicts a delay of about 26.71 percent (equal to 4.39 degree, see Table 1) for an airfoil with an actuator in the 5% position and a delay of 22.90 percent for an airfoil with an actuator in the LE. The actuator in the 25% of the delay has the smallest latency, which is 13.92 percent of the total delay. As a preliminary conclusion, at high AOAs, the optimal performance is related to the 5% actuator. The pressure drag is considerably decreased and the total drag is lowered with this actuation at high AOAs. This is because the separating zone and inverse flows have been removed in some degree.

Figure 7 depicts the percentage increase in the lift-to-drag ratio to the clean airfoil for the five AOAs of 18, 20, 22, 24, and 26 degrees, depending on the position of the actuators on the airfoil. At an AOA of 18 degrees, airfoils with actuators in 25% and 5% locations have the same and desired performance in terms of increasing lift to drag. Even the performance of an airfoil with an actuator in the 20% position outperforms both. The reason for this is because airfoils with actuators in the 20% and 25% positions have a bigger drag decrease. The performance of the actuated airfoil at the LE is also lower than that of the other airfoils in this diagram.

The impact of utilizing a plasma actuator in the 5% position at a 20-degree AOA is still greater than in other actuator installation positions. With the actuator at the 20%, the airfoil still performs well in this AOA. However, the drag of actuated airfoils in the 20 and 25 percent positions increases dramatically in the 22-degree AOA, in addition to reducing lift. As a result, with the AOA of 22 degrees, these two airfoils show a significant decline.

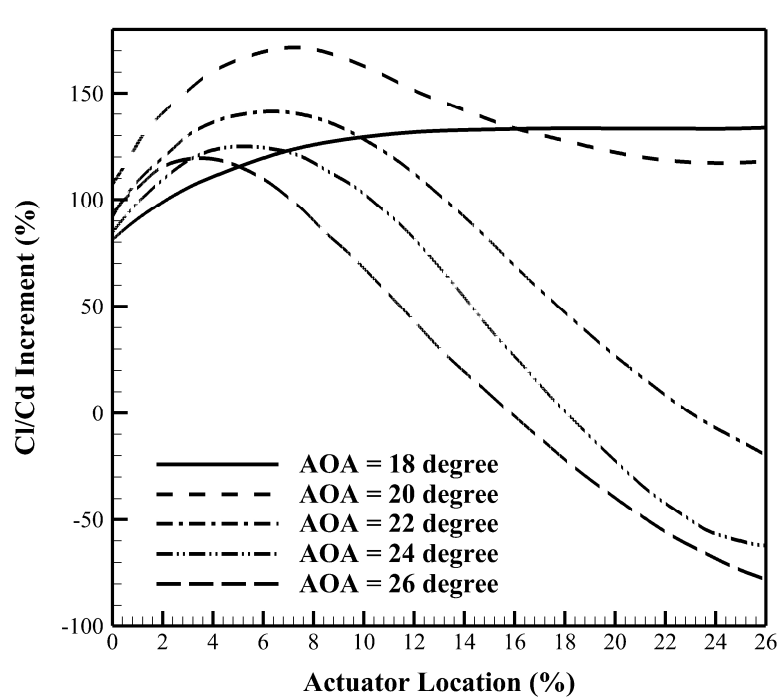

**Figure 7. Percentage increment in lift-to-drag ratio based on actuator location for AOAs of 18, 20, 22, 24, and 26 degrees (First Phase)**

At a 24 degree AOA, the poor performance of actuated airfoils at the 20 percent and 25 percent positions persists, as shown in Figure 7. Additionally, the 5 percent actuator is recognized as the best performance. In general, performance was reduced for all of the airfoils evaluated as compared to the lower AOAs. Also, at an AOA of 26 degrees, the performance of this airfoil diminishes due to the large reduction in drag given by the actuated airfoil in the 15% position. At this angle, the performance of the 15 percent actuated airfoil drops as well. There is also an improvement in the performance of the 10% and LE actuated airfoils.

At high AOAs, the regions between the LE and the 5 percent of the chord from the LE perform better. As a result, when the actuator is installed in these regions, it is feasible to achieve more desired aerodynamic coefficients. Thus, the DE optimization code was used in the second stage for the locations in this range. Four images of streamlines at a 24-degree AOA are shown in Figure 8. These four images show airfoils with plasma actuators at the leading edge, at 5%, 15%, and 25% of the chord from the leading edge, respectively. The performance of the airfoil with the actuator at the 5% location is likewise quite advantageous at this angle, as can be observed. The separation region entirely covers the actuator in an airfoil with an actuator in the 25% location, and the momentum injection in direction of the wall jet forms a vortex inside the separation area, which not only has no positive impact but also has a negative effect on the separated flow (see Figures 8 for 24 degree and Figure 9 for 26 degree of AOA).

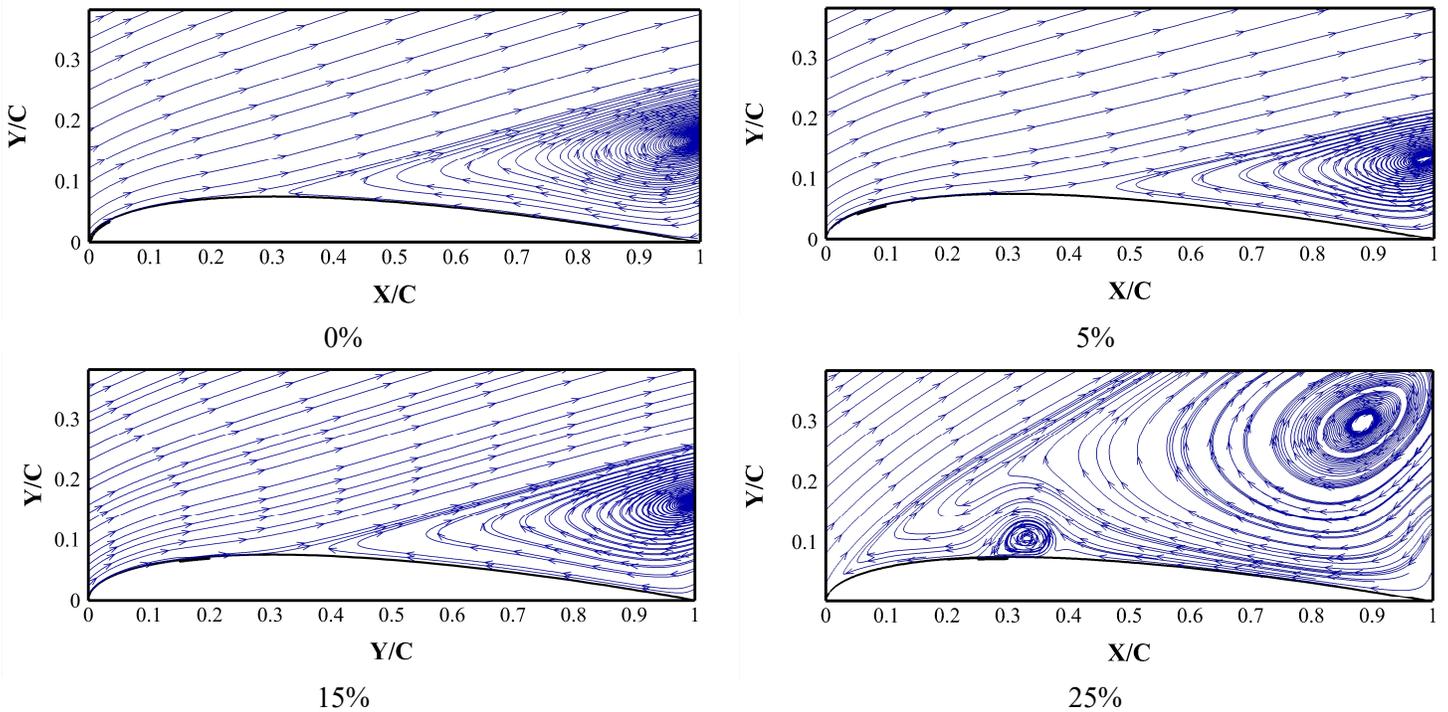

**Figure 8. The effect of plasma actuator locations on streamlines at 24-degrees AOA**

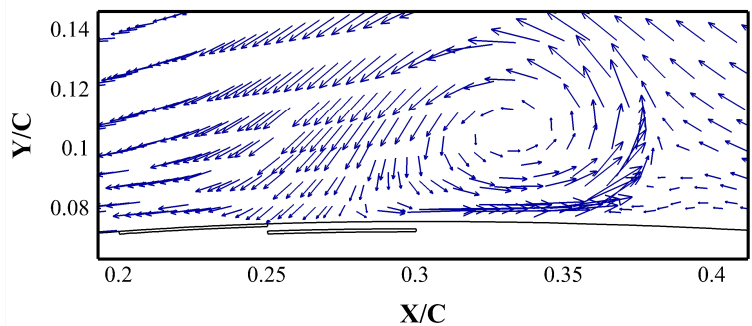

**Figure 9. The effect of plasma actuator 25% location on velocity vector at 26-degrees AOA**

## 5.2. Optimization Phase 2

For AOAs with various Reynolds numbers in the given range, the DE optimization approach evaluates locations between the LE and 5% of the chord length from the LE to find the best position for plasma actuator placement. As an example, the performance of airfoils with plasma actuators at AOAs of 20 and 22 degrees has been provided in Figure 10. For AOAs of 20 and 22 degrees, this figure illustrates the lift per drag coefficient increment for plasma-actuated airfoils in various locations. Reynolds numbers of 300, 600, and 900k are depicted on this graph. For AOA of 20 degrees, the best spots to mount the actuator are 4.248, 3.367, and 2.301 for Reynolds numbers of 300k, 600k, and 900k, respectively. Increase the AOA from 20 to 22 degrees notices this ideal value more clearly. At AOA of 22 degrees, the lift coefficient increases at a faster rate, specifically for lower Re numbers.

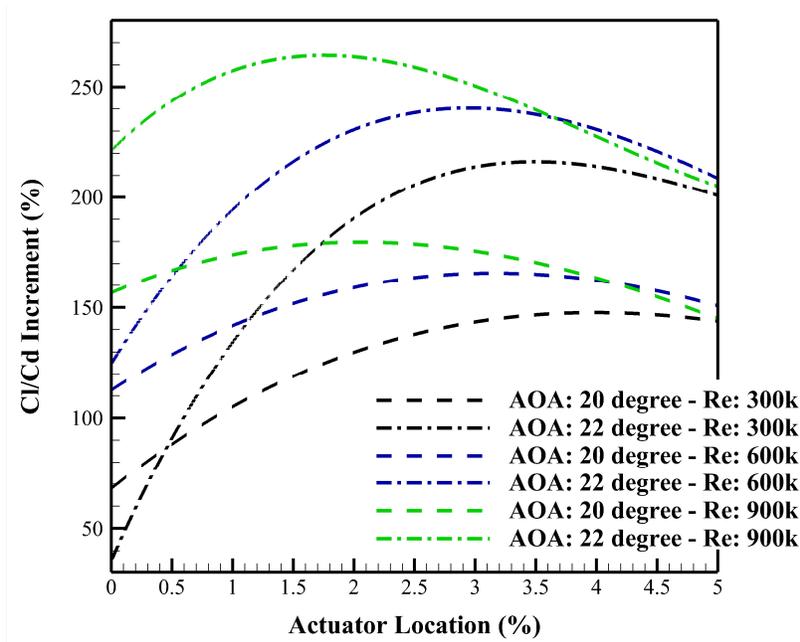

**Figure 10. Percentage increment in lift per drag coefficient based on actuator location for AOAs of 20 and 22 degrees, Blue: Re 300k, Black: Re 600k, Red: Re 900k**

Figure 11 depicts the optimization results for certain Reynolds numbers and AOAs. The horizontal axis of this map shows angle of attacks and the vertical axis is the Reynolds number × 0.001 and the optimum location of the plasma actuator as a percentage of the chord length from the leading edge is shown as a color points in this graph. According to the data, the best location for installation is between 2% and 4% of the chord from the leading edge of the airfoil. In a continuous optimization domain, a mathematical behavior for this data may be considered, and it can be applied at equivalent levels. According to the optimization data, the optimum location changes with Reynolds number and angle of attack, and may be characterized as a complicated function of a linear Reynolds effect and a power effect of the AOA (Equation (22)).

$$X_{Optimum} = ac^b$$
$$a = 28 - 0.02\frac{Re}{1000}$$
$$b = -0.58 + 0.0001\frac{Re}{1000}$$
$$c = AOA\ (^o)$$

(22)

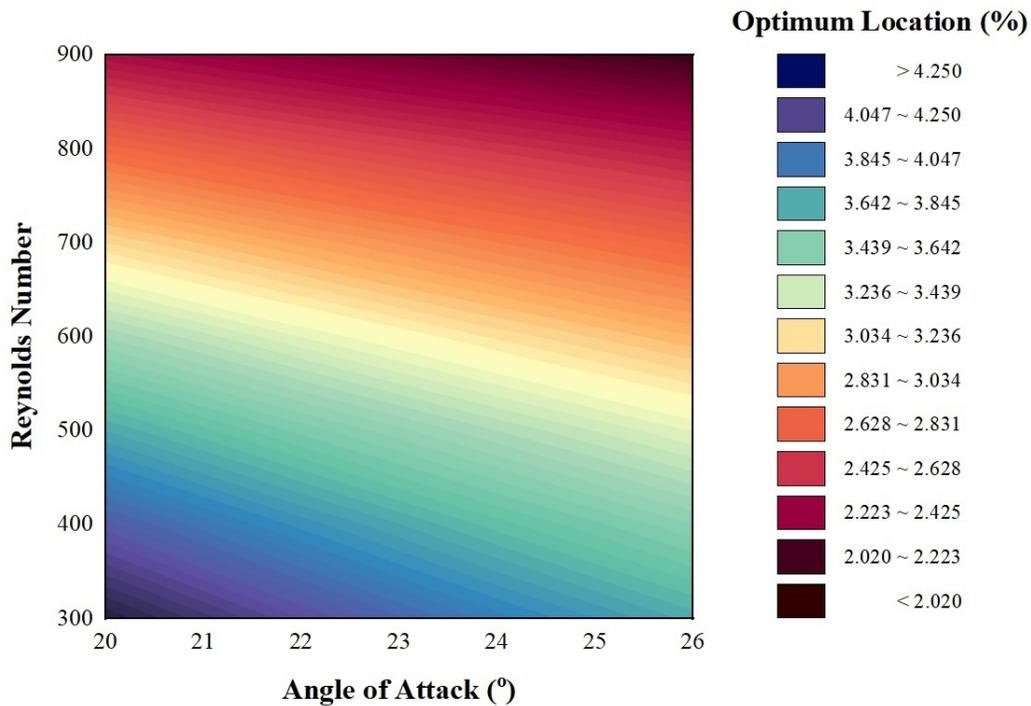

**Figure 11. Optimum plasma actuator locations regime map for various Reynolds number and angle of attacks**

# 6. Conclusion

This research uses Artificial Intelligence to optimize the placement of a Dielectric-Barrier-Discharge plasma actuator on a curved surface under varied aerodynamic and physical conditions. Differential Evolution (DE) was employed as the optimization technique for this work, and we used our previously explored improved electrostatic model to solve the electrostatic equations. We solved two elliptic equations of electrical potential and plasma density in this model, and the results were used as a source term in the momentum equation. The impact of plasma actuator installation location on the control of fluid flow separation from the NACA 0015 airfoil found that in AOAs near to the stall, actuators located 25 and 20 percent chord from the leading edge are optimum. The lift to drag ratio is relatively high in this region of the AOA because the drag coefficient of these installation positions is so low. This function is lost as the angle of attack increases, and higher performance is relocated to the point 5% of the chord length from the leading edge. The findings demonstrate that, despite previous research suggesting that the plasma actuator be installed on the airfoil's leading edge, the optimal location for plasma actuation is between 2% and 4% of the chord from the leading edge. A mathematical behavior is also taken into account for optimization data in a continuous domain, and it may be applied to any other geometry or design. Finally, the ideal site is given as a complicated function of a linear Reynolds effect and the angle of attack's power impact.